# Ion-Induced Dipole Interactions and Fragmentation Times : Cα-Cβ Chromophore Bond Dissociation Channel

Satchin Soorkia[a], Christophe Dehon[a], S. Sunil Kumar[a,*], Marie Pérot-Taillandier[a],

Bruno Lucas[a], Christophe Jouvet[b], Michel Barat[a] and Jacqueline A. Fayeton[a]

[a]Institut des Sciences Moléculaires d'Orsay, CNRS UMR 8214, Université Paris Sud, F-91405 Orsay Cedex, France

[b]Physique des Interactions Ioniques et Moléculaires (PIIM), UMR 7345, CNRS, Aix-Marseille Université, 13397 Marseille Cedex 20, France

**Corresponding Author**

satchin.soorkia@u-psud.fr

**Present address**

*Max-Planck-Institut für Kernphysik, Saupfercheckweg 1, 69117 Heidelberg, Germany

**ABSTRACT**

The fragmentation times corresponding to the loss of the chromophore (Cα-Cβ bond dissociation channel) after photo-excitation at 263 nm have been investigated for several small peptides containing tryptophan or tyrosine. For tryptophan-containing peptides, the aromatic chromophore is lost as an ionic fragment (*m/z* 130) and the fragmentation time increases with the mass of the neutral fragment. In contrast, for tyrosine-containing peptides the aromatic chromophore is always lost as a neutral fragment (mass = 107 amu) and the fragmentation time is found to be fast (< 20 ns). These different behaviours are explained by the role of the post-fragmentation interaction in the complex formed after the Cα-Cβ bond cleavage.

**TOC GRAPHICS**

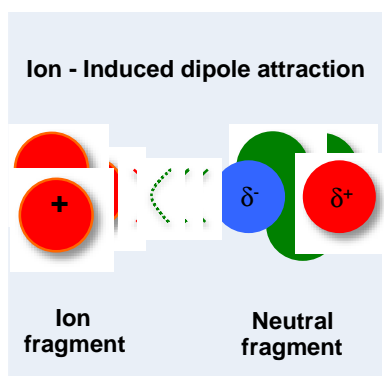

**KEYWORDS**

UV photodissociation, Peptides, Tryptophan, Tyrosine, Fragmentation times, Ion – induced dipole interactions

The Cα-Cβ bond breaking is a specific fragmentation process in small peptides containing tryptophan and tyrosine occurring after photo-excitation in the UV. Recently, new spectroscopic experiments and calculations on protonated tyrosine and phenylalanine have shown the extreme dependence of the UV induced fragmentation channels upon the electronic nature of the electronic excited state.[1] The photodissociation of protonated tryptophan and tyrosine have been studied at 263 nm using a multicoincidence approach.[2] For protonated tryptophan, it has been shown that *m/z* 130, resulting from Cα-Cβ bond dissociation, have two different origins and involve the coupling of the ππ* locally excited state with : (i) the ππ*$_{CO}$ state, which leads to a fast fragmentation occuring within τ < 20 ns after a proton transfer to the carbonyl group and (ii) the πσ*$_{NH3}$ state, which leads to a slow fragmentation with a characteristic τ > 1 ms, consecutive of a fast H-loss. In contrast, in the case of protonated tyrosine, only the fast fragmentation is observed. Note that the direct excitation of the πσ*$_{NH3}$ state in protonated tyrosine has been recently evidenced below 240 nm leading to the formation of the radical cation (H-loss channel), which subsequently fragments through Cα-Cβ bond dissociation forming the *m/z* 107.[3] Since the πσ*$_{NH3}$ charge transfer state is located lower in energy in protonated tryptophan, the excited state decays much more rapidly (380 fs) via H-loss and subsequent Cα-Cβ bond dissociation of the radical cation.[4]

Very recently, Zabuga et al.[5] reported a detailed study of the UV photofragmentation mechanism in Ac-FA$_5$K-H$^+$ and Ac-YA$_5$K-H$^+$, *i.e.* with phenylalanine and tyrosine as the UV chromophores, respectively. They proposed that intersystem crossing (ISC) plays a key role in the fragmentation of peptides, as evidenced by the large enhancement of the Cα-Cβ bond cleavage fragmentation channel by IR radiation up to tens of ms after the UV excitation. As for protonated tyrosine and phenylalnine, a picosecond pump-probe excitation scheme in a cold ion trap has been set up to record the excited state lifetimes as a function of the excess energy in the ππ* state.[6] The excited state lifetimes of the probed conformers decreases exponentially as a function of the excess energy imparted in the system with respect to the ππ* band origin. Moreover, these measurements are totally consistent with independent measurements using a femtosecond pump probe scheme at 266 nm at ambient temperature.[7] Since no constant signals at longer times (t > 1 ns) have been detected in the excited state lifetimes measurements for protonated tyrosine, it has been concluded that ISC to a triplet state is not at play for protonated tyrosine.

From an analytical point of view, undeniably, UV photodissociation, is complementary to other techniques such as Collisions-Induced Dissociation (CID)[8] or Infra-Red Multiphoton Dissociation (IRMPD)[9] for the study of peptides in mass spectrometry. At variance with IRMPD and CID, the use of UV photons allows deposition of a known amount of energy (4.71 eV at 263 nm) at a particular location (UV chromophore) in the peptide.[10] As a consequence, rapid[2] and specific fragmentations may occur such as loss of the side chain through Cα-Cβ bond cleavage as already demonstrated in several previous studies.[1,11,12] However, in conventional mass spectrometry, investigation of such bond ruptures relies solely on the detection of ion fragments after irradiation of the protonated species with limited information on the formation dynamics of the fragments.

In the case of a binary fragmentation (one ion and one neutral fragment), it can be argued that even if the Cα-Cβ bond is broken, the incipient fragments attract each other due to electrostatic forces and are stabilized in the form of an intermediate complex.[13] In a first approximation, these forces can be described in terms of ion-dipole and ion-induced dipole interactions,[14] as discussed further in this contribution. Moreover, for the same ion interacting with different neutral fragments, the ion-induced dipole force is expected to increase with increasing molecular weight, since the strength of the ion-induced dipole force is related to the polarizability of the neutral fragment.

In an attempt to bridge the gap between the protonated amino acid chromophores, *i.e.* protonated tyrosine and protonated tryptophan, and larger peptides, we have investigated the evolution of the fragmentation time of the Cα-Cβ bond cleavage reaction in a series of small peptides with increasing molecular weight containing tyrosine or tryptophan as the UV chromophore. Our unique multicoincidence experimental approach to UV photodissociation[15-20] ascertains that the Cα-Cβ rupture is indeed a binary fragmentation, allowing a systematic study of the evolution of the fragmentation time as a function of increasing molecular weight. For these systems, the ejected aromatic chromophores are respectively neutral for molecules involving tyrosine (mass = 107 amu) and ionic (*m/z* 130) for those involving tryptophan. This result is in accord with the higher ionization potential (DFT B3LYP/aug-cc-pVDZ calculations) of the side chain of tyrosine as compared with that of tryptophan.(private communications) The role of the post-fragmentation interactions between the ion and the neutral fragment on the dynamics of their seperation will be discussed.

Summarized in Table 1 and Table 2 are the fragmentation time (τ±Δτ in ns), the mass and the calculated polarisability of the neutral fragment for tryptophan- and tyrosine-containning peptides. Details on the determination of the fragmentation times, which is in essence the decay time of the distribution of the ion fragment (or neutral fragment) as a function of the time-of-flight, are given in the experimental method section and references therein. An example of such a decay measurement is given in Figure 1 for protonated tryptophan-methionine, where the neutral fragment has a mass = 206 amu. For the sake of discussion, we computed the polarizabilities, $\alpha$, of the neutral fragments using the additivity method for molecular polarizability by Miller.[21]

**Table 1 : Studied tryptophan-containing protonated peptides**

| Name | Letter | Mass of the parent ion | Fragmentation time $\tau\pm\Delta\tau$ (ns) | Mass of neutral fragment | Polarisability ($Å^3$) |
|---|---|---|---|---|---|
| Tryptophan | W | 205 | < 20 | 75 | 7.35 |
| Glycine-Tryptophan | GW | 262 | 31±5 | 132 | 12.8 |
| Tryptophan-Glycine | WG | 262 | 30±5 | 132 | 12.8 |
| Tryptophan-Cystine | WC | 308 | 213±20 | 178 | 16.76 |
| Glycine-Tryptophan-Glycine | GWG | 319 | 80±20 | 189 | 18.28 |
| Tryptophan-Methionine | WM | 336 | 120±10 | 206 | 21.8 |
| Methionine-Tryptophan | MW | 336 | 145±l5 | 206 | 21.5 |
| Tryptophan-Methionine(oxydized) | $WM_{ox}$ | 352 | 164±20 | 222 | 22.9 |
| Glycine-Tryptophan-Methionine | GWM | 409 | 200±20 | 279 | 24.9 |

**Table 2 : Studied tyrosine-containing protonated peptides**

| Name | Letter | Mass of the parent ion | Fragmentation time $\tau\pm\Delta\tau$ (ns) | Mass of neutral fragment | Polarisability ($Å^3$) |
|---|---|---|---|---|---|
| Tyrosine | Y | 182 | < 20 | 75 | 7.35 |
| Tyrosine-Glycine | YG | 239 | 33±10 | 107 | 13.6 |
| Tyrosine-Alanine | YA | 253 | < 20 | 107 | 13.6 |
| Tyrosine-Cystine | YC | 285 | 135±10 | 178 | 16.76 |
| Tyrosine-Lysine | YK | 310 | 36±10 | 107 | 13.6 |
| Lysine-Tyrosine | KY | 310 | < 20 | 107 | 13.6 |
| Methionine-Tyrosine | MY | 313 | < 20 | 107 | 13.6 |
| Tyrosine-Methionine | YM | 313 | < 20 | 107 | 13.6 |
| Tyrosine-Methionine(oxydized) | $YM_{ox}$ | 329 | < 20 | 107 | 13.6 |
| Methionine(oxydized)-Tyrosine | $M_{ox}Y$ | 329 | < 20 | 107 | 13.6 |
| Tyrosine-Lysine(acetylated) | $YK_{ac}$ | 352 | < 20 | 107 | 13.6 |
| Glycine-Tyrosine-Methionine | GYM | 370 | < 20 | 107 | 13.6 |
| Glycine-Tyrosine-Methionine(oxydized) | $GYM_{ox}$ | 386 | < 20 | 107 | 13.6 |

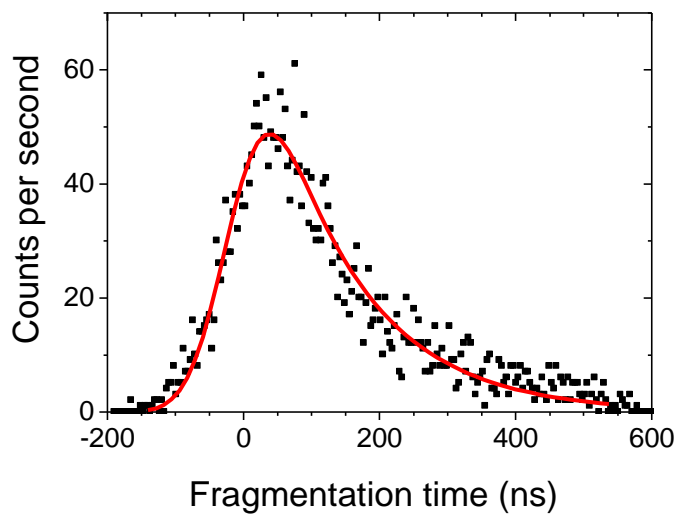

**Figure 1 : Fragmentation time distribution for the m/z 130 fragment of protonated tryptophan-methionine. Full line is a fit of the data with a Gaussian function (ω = 46±5 ns) convoluted with an exponential decay function giving τ = 120±10 ns**

The evolution of the fragmentation time distributions presents very different behaviours for molecules with a tryptophan or a tyrosine UV chromophore. For tryptophan-containing peptides the fragmentation time increases with the molecular mass of the parent molecule. This behavior is manifest in Figure 2(a), which depicts an increasing fragmentation time as a function of the mass of the peptide, with the exception of protonated Tryptophan-Cysteine (WC), which shows a fragmentation time that is heavily deviated from the smooth behavior of all other values. In contrast, for tyrosine-containing peptides, the fragmentation times are very short (< 40 ns) for all the studied species, with the exception of protonated Tyrosine-Cysteine (YC). Overall, the measure of the fragmentation time is seemingly independent of the molecular weight of the parent ion and comparable with the ultimate time resolution of the experiment (20 ns).

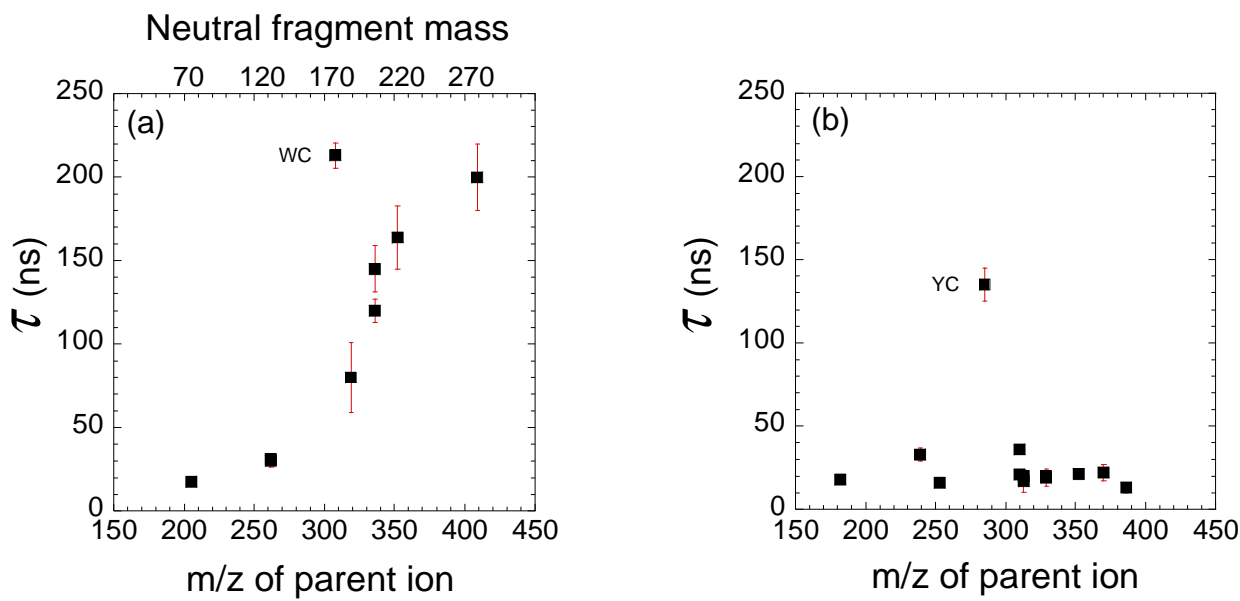

Figure 2 : Evolution of the Cα-Cβ fragmentation time τ(ns) as a function of the mass of the parent ion for (a) tryptophan-containing peptides and (b) tyrosine-containing peptides. Note that the scale on top of (a) is for the mass of the neutral fragment and no such scale is shown in (b) since the mass of the neutral fragment is equal to 107 (with the exception of YC).

We now focus on the marked difference in the trends of the fragmentation times corresponding to the loss of the UV chromophores. A comprehensive understanding of the fragmentation mechanism leading to the production of a neutral side chain loss for tyrosine-containing peptides (with the exception of YC) and an ionic side chain loss for tryptophan-containing peptides (with the exception of WC) deserves excited-state calculations, which are outside the scope of this paper. Again, the experimental multicoincidence approach is central to ascertain that the Cα-Cβ bond cleavage is indeed a direct (one ion and one neutral fragment) dissociation for both the tryptophan- and tyrosine-containing peptides. We recall that Cα-Cβ bond cleavage is specific to UV excitation.

While the fragmentation time increases with the mass of the neutral fragment for tryptophan-containing peptides, it is fast for tyrosine-containing peptides. These different behaviours could be due to the post-fragmentation ion-induced dipole interaction in the complex formed after the Cα–Cβ bond cleavage as detailed hereafter. The major difference lies in the size of the neutral fragment which is released in the fragmentation process. On the one hand, for tyrosine-containing peptides, the neutral fragment is always the same, *i.e.* mass = 107 amu, while the ion fragment has an increasing *m/z*. On the other hand, for tryptophan-containing peptides, the ion fragment is always the same, *i.e. m/z* 130, while the neutral fragment has an increasing molecular mass. After Cα-Cβ bond rupture, it is assumed that the ion and the neutral fragment are bound together in the form of a loose complex. The stabilization energy V(r) of such a complex may be approximated by equation 1.[14]

$$V(r) = -\frac{\mu_d e \cos\theta}{4\pi\varepsilon_0 r^2} - \frac{\alpha e^2}{8\pi\varepsilon_0 r^4} \quad (1)$$

The first term represents the contribution of the permanent dipole moment of the neutral ($\mu_d$ is the dipole moment, $e$ is the electronic charge, $\theta$ is the angle between the direction of the dipole and the charge-dipole axis, $\varepsilon_0$ is the vacuum permittivity and $r$ is the distance between the center of charge of the ion and the midpoint of the dipole). The second term represents the contribution of the interaction between the ion and the induced dipole moment of the neutral ($\alpha$ is the molecular polarizability). In a peptide, the main contribution to the electric dipole is due to the peptide bond between amino acids (~3.5 D per bond).[22] Since the fragmentation times measured are on the order of tens to hundreds of nanoseconds, it can be argued that the mean value of the cosine function is zero. This stems from the assumption that in the case of a loose complex, the ion and the neutral fragment rotate freely, *i.e.* there is no prefered orientation of the ion with respect to the dipole, on this timescale. The second term, which is proportional to the polarizability, has to be overcome in order for the incipient fragments to separate.

The polarizability (α in Å³) as a function of the mass of the neutral fragment are shown in Figure 3. Clearly, α increases (linear trend) with the mass of the neutral fragment, which indicates that the ion-induced dipole interaction increases as the mass of the neutral fragment increases in the ion-neutral complex.

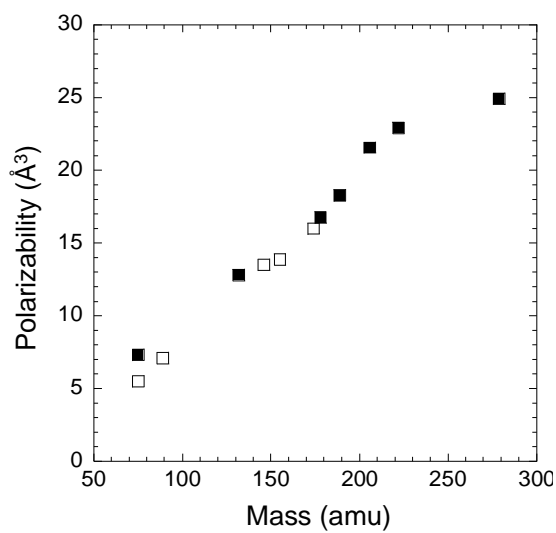

**Figure 3** Polarizability α of the neutral fragments as a function of their masses : full square values are obtained with the Miller additiviy method for molecular polarizability and open square values are calculated values from ref. 23.

Furthermore, assuming that the rate constant ($k = 1/\tau$) of the fragmentation reaction at a given temperature T is related to the activation energy $E$ by the Arrhenius equation given in equation (2),

$$k = \frac{1}{\tau} = Ae^{\frac{-E}{RT}} \qquad (2)$$

and : (i) the temperature of all studied species is the same as that of the electrospray source, (ii) the activation energy is proportional to the polarizability of the neutral fragment and (iii) the internal energy is the same after absorption of the UV photon and dissociation of the Cα-Cβ bond in the first approximation for all compounds, one can estimate crudely the average ion-neutral distance at which the two fragments interact. Shown in Figure 4 is a plot of the fragmentation time as a function of the polarizability of the neutral fragments. The red line is a fit of the data points with a monoexponential function. As can be seen, the data points (with the exception of protonated WC and YC as discussed later in this section) follow a linear evolution (ln y scale) as a function of α.

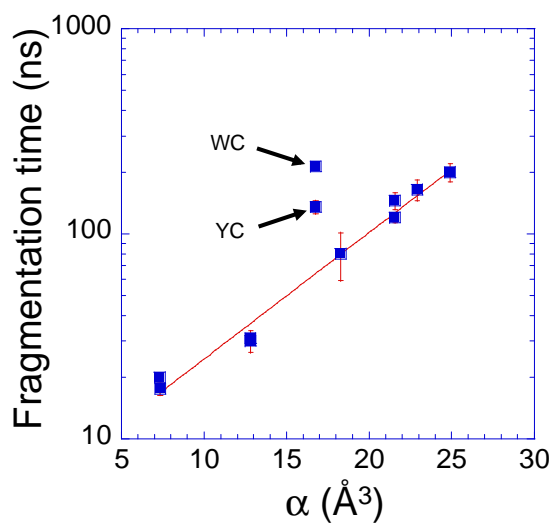

**Figure 4 : Plot of the fragmentation time (τ/ns) as a function of the polarizability of the neutral fragments for tryptophan-containing peptides. The fragmentation times of protonated WC and YC, which deviate from the general trend, are also indicated on the plot (see text). All data points, with the exception of protonated WC and YC are fitted with a monoexponential function (red line).**

With an effective temperature of 355K,[5] the average ion-neutral distance amounts to 6.4 Å, which is consistant with the distance between the center of the indole chromophore and the Cα carbon atom in protonated tryptophan.[24]

As can be seen in Figure 2, protonated WC and YC do not follow the same trend as the other species. In fact, the velocity vector (vv) correlation diagram[15] for WC does not correspond to a binary fragmentation. The vv correlation (see supplementary information) suggests the superposition of two structures : (i) a first structure which has a slope very close to that of a binary -178/130 fragmentation (within experimental uncertainties) and (ii) a second structure with a slope almost equal to zero, corresponding to the ejection of a very light neutral fragment. Actually, the detailed analysis of the data clearly demonstrates triple coincidence events. By analogy with protonated tryptophan, *m/z* 130 in protonated WC could result from a two-step mechanism, *i.e.* a fast H loss prior to Cα-Cβ bond cleavage. Unfortunately, the fragmentations times of the two steps are too close for a precise determination of the mechanism. Nevertheless, it is interesting to note that the fragmentation time ($\tau = 213\pm20$ ns), though slower than a direct Cα-Cβ bond rupture, is even faster than the overall fragmentation time of the equivalent slow fragmentation pathway ($\tau > 1$ ms) leading to *m/z* 130 ions in protonated tryptophan.[2]

With regard to YC for which the measured fragmentation time ($\tau = 135\pm10$ ns) is almost a factor of three larger than the other species, the side chain is lost as an ionic fragment (*m/z* 285 → *m/z* 107 + 178). Hence, this species cannot be directly compared to the other tyrosine-containing species. Moreover, this value is significantly close to the one that can be predicted from the plot of Figure 2(a) for a species with a *m/z* 285.

In conclusion, we report here on the evolution of the Cα-Cβ fragmentation time measured in a series of tyrosine- or tryptophan-containing small peptides. The multicoincidence approach to UV photodissociation allows a systematic study of the latter UV specific fragmentation channel, and in particular the loss of the UV chromophore in a binary fragmentation. After excitation at 263 nm, the aromatic chromophore is released as an ion fragment (*m/z* 130) for tryptophan-containing peptides and the fragmentation time is found to increase as a function of the mass of the neutral fragment. This experimental observation can be interpreted in terms of ion-induced dipole interactions between the incipient fragments. Conversely, the charge is always localized on the backbone for tyrosine-containing peptides, *i.e.* loss of the UV chromophore as a neutral fragment of mass = 107 amu, and the fragmentation times (within the experimental limits and errors) are found to be independent of the size of the system.

**EXPERIMENTAL METHOD**

The experimental setup has been described in detail elsewhere[15-17] and only a brief description is presented here. All dipeptides used in this study are obtained commercially (GeneCust, 95% purity). Protonated molecules, produced in an electrospray ion source as ion bunches of ~130 ns duration are mass-selected and accelerated to 5 keV. The pulsed ion beam intercepts the 263 nm laser pulse (100 µJ/pulse, 200 ns duration and 1 kHz pulse repetition rate) in the interaction region. The interaction box is made of a set of electrodes that can be polarized to create an electric field oriented along the beam axis. This allows determining the fragmentation times in a range from a few tens up to hundreds of nanoseconds for each fragmentation channel as detailed elsewhere.[17] An event is defined as a coincidence between an ion and a neutral fragment generated in the dissociation of a given ion during one laser pulse. We have ascertained that the C$\alpha$-C$\beta$ ruptures are binary fragmentations as discussed elsewhere.[16] For each event, the fragmentation time is given by the difference between the arrival times of the ion ($t_i$) and of the neutral ($t_n$) fragments on their respective detectors.[16] Thus we are not limited by the pulse width of the laser beam. Nevertheless, the time spread due to the kinetic energy release (KER) cannot be compensated. The only recorded events are those with a fragmentation time smaller than 5 µs, the time spent by the ions between the interaction zone and the entrance of the electrostatic analyser. The time dependence of the fragmentation rate is then derived from the ($t_i - t_n$) distribution given by a linear relationship.[16] The fragmentation time is then obtained from a fit of this curve by the convolution of a gaussian function, which accounts for the kinetic energy release distribution and of an exponentially decreasing function for the fragmentation decay time.

**ACKNOWLEDGEMENTS**

The present work was carried out at the "Centre de Cinétique Rapide ELYSE". C.D. and S.K. are thankful for financial support provided by Université Paris-Sud XI, CNRS and the ANR Research Grant (ANR 2010 BLANC 040501), for carrying out this work. The authors thank Gilles Grégoire and Michel Broquier for stimulus and fruitful discussions.